\documentclass{aastex}
\usepackage{psfig}

\def\asca{{\it ASCA\/}}
\def\chandra{{\it Chandra\/}}

\def\rxte{{\it RXTE\/}}

\def\cir{Cir~X-1}

\def\ltsima{$\; \buildrel < \over \sim \;$}
\def\simlt{\lower.5ex\hbox{\ltsima}}
\def\gtsima{$\; \buildrel > \over \sim \;$}
\def\simgt{\lower.5ex\hbox{\gtsima}}

\slugcomment{Accepted for publication in The Astrophysical Journal Letters}

\shorttitle{X-RAY P~CYGNI LINES FROM CIRCINUS~X-1}

\shortauthors{BRANDT \& SCHULZ}

\begin{document}

\title{The Discovery of Broad P~Cygni X-ray Lines from Circinus~X-1\\
with the {\it CHANDRA\/} High Energy Transmission Grating Spectrometer}
\author{
W. N. Brandt\altaffilmark{1} 
and
N. S. Schulz\altaffilmark{2}
 }
\altaffiltext{1}{Department of Astronomy and Astrophysics, 525 Davey
Laboratory, The Pennsylvania State University, University Park, PA, 16802.}
\altaffiltext{2}{Center for Space Research, Massachusetts Institute of Technology, 
Cambridge, MA 02139.}
\begin{abstract}
We present the first grating-resolution X-ray spectra of the X-ray binary
\cir, obtained with the High Energy Transmission Grating Spectrometer on 
\chandra. These reveal a rich set of lines from 
H-like and/or He-like Ne, Mg, Si, S and Fe detected with a high 
signal-to-noise ratio. The lines are 
broad ($\pm 2000$~km~s$^{-1}$) and show P~Cygni profiles. The absorption 
components of the lines extend to low velocity, and they have about the 
same widths and strengths as the corresponding emission components. The 
widths of the X-ray P~Cygni lines are comparable to that of the broad 
component of the strong, asymmetric H$\alpha$ line from \cir, 
suggesting that the two phenomena may be related. We discuss 
outflow models and propose that the P~Cygni profiles may 
arise in the moderate temperature ($\sim 5\times 10^6$~K) region of 
the wind from an X-ray heated accretion disk. This basic
picture strengthens the idea that the accretion disk in \cir\ is
viewed in a relatively edge-on manner, and it suggests that \cir\ 
is the X-ray binary analog of a Broad Absorption Line quasar. 
\end{abstract}


% --------------------------------------------------------------------------------------

\section{Introduction}
\label{intro}

The most sensitive indicators of mass loss from cosmic sources are the
resonance absorption lines of abundant ions. In the ultraviolet, for example, 
these lines have been used to study outflows in objects ranging from 
stars to quasars. Prior to the launch of the 
\chandra\ X-ray Observatory (e.g., Weisskopf et~al. 2000), 
however, X-ray spectrometers generally lacked the resolution and
sensitivity needed to study resonance line absorption by the outflows 
from most cosmic objects. Resonance absorption line studies
in the X-ray band offer a number of attractive features: 
(1) the continuum driving the ionization is often directly visible, 
(2) the atomic physics of the relevant ions is often relatively simple, and
(3) unlike ultraviolet photons, X-rays are not easily destroyed 
in dusty environments. 

The mysterious X-ray binary \cir\ is thought to contain a neutron star 
that can radiate at up to super-Eddington levels near the periastron
passage (`zero phase') of its 16.5-day orbital cycle 
(e.g., Tennant, Fabian \& Shafer 1986; Inoue 1989), and its accretion
disk is probably viewed in a relatively edge-on manner (e.g., Brandt et~al. 1996). 
The fact that it can radiate with high luminosity relative to its Eddington 
luminosity ($L/L_{\rm Edd}$) makes it a natural system in which to expect 
observable outflows, due to the larger amount of photon pressure available
per unit gravitational mass. Indeed, the 
broad component (up to 2000~km~s$^{-1}$ FWHM) of the strong, 
asymmetric H$\alpha$ line from \cir\ has been interpreted as arising in an 
outflow from regions near the neutron star (e.g., Whelan et~al. 1977; 
Johnston, Fender \& Wu 1999); the broad component of the H$\alpha$ line
is blueshifted relative to the narrow component. 
Furthermore, the system appears to emit
radio jets and is embedded in a synchrotron nebula (e.g., Stewart et~al. 1993; 
Fender et~al. 1998). Finally, we note a convergence of opinion that the 
system is probably a low-mass X-ray binary based on 
photometry and variability of the optical counterpart (e.g., Stewart et~al. 1991; Glass 1994), 
its orbital period and apparent space velocity (Brandt \& Podsiadlowski 1996; Tauris et~al. 1999), 
its correlated X-ray spectral and timing properties (e.g., Shirey, Bradt \& Levine 1999), 
and the observed type~1 X-ray bursts (Tennant et~al. 1986).

\nopagebreak[3]

% This Letter describes the discovery of broad lines in the X-ray
% spectrum of \cir\ that show P~Cygni profiles. In \S2 we describe
% the observations and data analysis, and in \S3 we give a first
% discussion of the results. To our knowledge, the only previous
% observation of \cir\ with an X-ray grating spectrometer was with
% the Objective Grating Spectrometer (OGS) on \einstein\ 
% (Vrtilek et~al. 1991); \cir\ was in a low state during
% the OGS observation and was not detected. 

% --------------------------------------------------------------------------------------

\section{Observations and Data Analysis}
\label{obser}

\cir\ was observed using the High Energy Transmission Grating Spectrometer 
(HETGS; C.R. Canizares et al., in preparation) on \chandra\ starting at
22:09:50 UT on 2000 February 29 (see Figure~1).\footnote{For additional
information on the HETGS and ACIS see the \chandra\ Proposers' Observatory 
Guide at http://asc.harvard.edu/udocs/docs.} The total integration time 
was 32~ks, and the observation was continuous. The observation start time 
corresponds to phase 0.99 using the ephemeris discussed by Glass (1994). 

The HETGS carries two types of transmission gratings, the 
Medium Energy Grating (MEG) and the High Energy Grating (HEG). It allows 
high-resolution spectroscopy from $\approx$~1.2--31~\AA\ ($\approx$~0.4--10~keV) 
with a peak spectral resolution at 
$\approx 12$~\AA\ of $\lambda/\Delta\lambda \approx 1000$ for HEG
1st order. The resolution for higher orders improves by a factor of $n$ for the 
$n$th order, but the spectral bandpass and efficiency are reduced for these
orders. In this analysis we mostly utilize the MEG and HEG 1st order as well 
as the MEG 3rd order spectra. 

The dispersed spectra were recorded with an array of 6 Charge Coupled 
Devices (CCDs) which are part of the Advanced CCD Imaging 
Spectrometer (ACIS; G.P. Garmire et~al., in preparation).$^3$ To avoid CCD 
frame `dropouts' due to telemetry saturation, we blocked the 0th order 
image with a `spatial window' and shut off the two peripheral CCDs (S0 and S5). 
The omission of the peripheral CCDs could be tolerated because \cir\ is heavily 
absorbed by the interstellar column density 
($N_{\rm H}\approx 2\times 10^{22}$~cm$^{-2}$; e.g., 
Predehl \& Schmitt 1995) and offers hardly any flux 
above $\approx 15$~\AA\ (e.g., Morrison \& McCammon 1983). 
The brightness of \cir\ also required additional mitigation 
efforts for `photon pile-up' effects. We applied a `subarray' during the 
observation that reduced the CCD frame time to 1.7~s. The remaining pile-up 
effects caused the 1st order spectra below $\approx 10$~\AA\ (MEG) and 
$\approx 6$~\AA\ (HEG) to be depleted of photons, which then populate the 
higher order spectra at shorter wavelengths. Below 7~\AA\ the MEG 1st order 
spectrum is so depleted that we will not use it further in this analysis. 
The MEG 3rd order spectrum is pile-up free from 2.3--9~\AA\ (1.4--5.4~keV). 
Most of the lines we present below are not or are only moderately affected by
pile-up. 

Due to a misalignment of the applied subarray during the observation, 
only half of the dispersed spectra were fully recorded; we only have 
negative orders in the MEG and positive orders in the HEG complete. 
The \chandra\ X-ray Center (CXC) provided aspect-corrected level~1 event 
lists via its standard pipeline processing. 

% We reprocessed these data using the latest available data processing input 
% products. We also removed events that resulted from either bad pixels or 
% columns as well as from flaring background.

The determination of the 0th order image position is crucial for the 
calibration of the wavelength scales because it defines their zero 
point. We calculated this position by fitting the dispersed images of the 
MEG and HEG and by determining the intersections of these fits with the 
0th order `read-out trace' on the S3 CCD. The fits of the MEG and HEG were 
consistent to within 0.3 detector pixels, which translates into a zero point 
accuracy of 0.004~\AA\ for the MEG \,$-1$st order and 0.002~\AA\ for the 
HEG \,$+1$st order. The overall wavelength calibration is good to $\approx 0.1$\% 
and mostly depends on uncertainties in the positions of the CCD chip 
gaps, which currently are $\pm 0.5$ pixels. For most of the spectral features
analyzed here there is only one gap involved leaving us with an overall 
worst case uncertainty in the scale of 
0.008~\AA\ for MEG \,$-1$st order, 
0.005~\AA\ for HEG \,$+1$st order, and 
0.003~\AA\ for MEG \,$-3$rd order. 

We then processed the events into final event lists using CXC software, and
we used custom software and {\sc ftools} to produce our final grating spectra. 
After standard grade selection, we have a total of 
$1.35\times 10^6$ events in MEG \,$-1$st order, 
$1.52\times 10^6$ events in HEG \,$+1$st order, and 
$3.84\times 10^5$ events in MEG \,$-3$rd order. 
We corrected our final grating spectra for instrumental effects 
(e.g., energy dependent photon detection efficiency and gaps between the CCDs) 
using aspect-corrected exposure maps. 

From the MEG \,$-3$rd order spectrum we determined a mean 2--8~keV flux of 
$1.8\times 10^{-8}$~erg~cm$^{-2}$~s$^{-1}$. This was calculated by fitting 
the \asca\ continuum model of Brandt et~al. (1996) to the 2.3--6.2~\AA\ 
(2--5.4~keV) spectrum (where pile-up is unimportant) and then extrapolating
the result to the 2--8~keV band. The observed flux is 
within the range that has recently been seen by
\rxte\ (Shirey et~al. 1999) at zero phase. Even for a fairly small distance
of $\approx 6$~kpc (see \S3 of Stewart et~al. 1993 and 
Case \& Bhattacharya 1998), the 2--8~keV luminosity is 
$L_{2-8}\simgt 9.6\times 10^{37}$~erg~s$^{-1}$ (corrected for interstellar
absorption); hence for a $M=1.4$~$M_\odot$ neutron star with
$L_{\rm Edd}=1.3\times 10^{38} (M/M_\odot)$~erg~s$^{-1}$ we find
$L_{2-8}/L_{\rm Edd}\simgt 0.5$ (note there is certainly substantial 
X-ray luminosity below 2~keV as well). 
Figure~2 shows the resulting HEG \,$+1$st and MEG \,$-1$st order spectra. These 
reveal a rich set of highly significant lines from H-like and/or He-like 
Ne, Mg, Si, S and Fe. The lines are broad ($\pm 2000$~km~s$^{-1}$) and 
show P~Cygni profiles (see Figure~3).

Table~1 gives a list of the lines identified with P~Cygni profiles; many 
of these have two or more independent detections (see Figure~3 for an 
example). We used the line list presented in 
Mewe, Gronenschild \& van den Oord (1985) to determine the line 
wavelengths and transition types. Most measured wavelengths agree 
to within 0.01~\AA\ with the predicted wavelengths. Some wavelengths
are measured to be slightly lower, which may be attributed to the fact 
that some of the profiles are broadened and not entirely symmetric. 
The `peak-to-valley' distances for the P~Cygni profiles are a 
measure of the broadening of the lines. Table~1 shows that the 
widths are quite significant (0.02--0.10~\AA), well above the 
resolution of the HETGS.  
%
% We have divided our observation into several segments and examined
% spectra for each of these. 
%
While there is evidence for line variability over time, P~Cygni 
profiles are persistently present. 

% visible in all spectra. 

% --------------------------------------------------------------------------------------

\section{Discussion}

% \subsection{Absorption Lines, Column Density and Ionization Parameter}
% \label{ionsec}

The observed X-ray P~Cygni profiles are most naturally interpreted as arising 
in a high-velocity outflow from the \cir\ system. Several properties of 
the profiles deserve note. 
First of all, the absorption troughs extend to low velocities; this 
suggests that we are seeing the region in which the outflow is accelerated, 
rather than, for example, just a shell of material that has already attained 
its terminal velocity. 
Secondly, the strengths of the emission and absorption components are 
comparable when integrated over the whole observation. 
Finally, the best-defined profiles look fairly smooth and continuous
down to the resolution limit of the HETGS; at least down to this limit
there is no evidence for strong shell-like density enhancements or 
other inhomogeneities in the outflow. 
The size of the \cir\ system is uncertain but is thought to be
$\sim 5\times 10^7$~km (e.g., Tauris et~al. 1999), so the characteristic 
time for the outflow to cross the system is $\sim 0.5$~days. 

As mentioned in \S1, an outflow from the \cir\ system has already been
invoked to explain the broad component of its H$\alpha$
line (Johnston et~al. 1999). The widths of the X-ray P~Cygni profiles are 
quite consistent with the FWHM of the broad component of H$\alpha$, 
suggesting that these two phenomena may be related; ultraviolet P~Cygni 
profiles and optical H$\alpha$ emission are often seen together from stars 
with high mass loss rates (see \S2.3 of Lamers \& Cassinelli 1999). 
Johnston et~al. (1999) argued that the broad H$\alpha$ line component
arises in a strong, anisotropic outflow driven by the radiation pressure
from the neutron star. In their model, this outflow must be optically 
thick (to H$\alpha$ photons) to explain the broad blue wing of the 
line (the redshifted emission being obscured by the outflow itself). 

The observed P~Cygni profiles are broadly consistent with such an outflow 
model, although this outflow must be fairly dense to avoid becoming 
completely ionized by the large X-ray luminosity of \cir. 
In particular, the ionization parameter $\xi=L/nr^2$ must be 
$\simlt 1000$~erg~cm~s$^{-1}$ so that the ions listed in 
Table~1 can survive (see Kallman \& McCray 1982).\footnote{In
the definition of $\xi$, $L$ is the ionizing luminosity, 
$n$ is the electron number density, and 
$r$ is the distance to the ionizing radiation source.}
If we take the launching radius (i.e., onset radius)
of the outflow to be $r_{\rm launch}\approx 10^5$~km,
so that its observed terminal velocity, $v_{\rm t}$, 
is comparable to the escape velocity, $v_{\rm esc}$,
at $r_{\rm launch}$, we require $n_{\rm launch}\simgt 10^{15}$~cm$^{-3}$
(we consider the plausible case of outflows launched from larger
radii where $v_{\rm t}>v_{\rm esc}$ below).  
For comparison, the expected density of the accretion disk at this 
radius is $\approx 2\times 10^{18}$~cm$^{-3}$, so the outflow-to-disk 
density contrast seems plausible (compare with Figure~2 of 
Raymond 1993).\footnote{We have calculated the
disk density following \S5.6 of Frank, King \& Raine (1992) with
a neutron star mass of $M=1.4$~$M_\odot$, 
a mass accretion rate of $\dot M = 1.5\times 10^{18}$~gm~s$^{-1}$, and
a viscosity parameter of $\alpha=0.1$.}
To allow the survival of H~{\sc i} (for H$\alpha$) 
in the outflow, $\xi\simlt 1$~erg~cm~s$^{-1}$ and hence 
$n_{\rm H~I}\simgt 10^{18}$~cm$^{-3}$ is required; such a high density
outflow would be surprising, so it seems likely that most of the 
H$\alpha$ emission is formed at a larger distance in the outflow
or in the accretion disk (see Mignani, Caraveo \& Bignami 1997). 

Based on the X-ray spectral variability properties of \cir, 
Brandt et~al. (1996) argued that its accretion disk is flared and
viewed in a relatively edge-on manner. X-rays emitted in the inner
part of such a system can heat the surface of the disk farther
out, producing a corona and thermally driven wind 
(e.g., Begelman, McKee \& Shields 1983). The outflow revealed 
by our X-ray P~Cygni profiles might well be this wind, although 
for this high $L/L_{\rm Edd}$ system radiation pressure acting
on lines (in addition to Compton heating) is likely to be important. 
Raymond (1993) has examined 
atomic processes in the context of this scenario, and his 
calculations show that at distances of $\approx 10^5$--$10^6$~km 
there is a significant (subtending $\sim 3$--20$^\circ$), moderate 
temperature ($\sim 5\times 10^6$~K) region where atomic heating and 
cooling processes dominate over Compton processes. Significant line 
emission is expected from this region, and in fact most of the lines 
we list in Table~1 are those predicted to be strong by Raymond (1993). 
Thus, our favored interpretation for the observed X-ray P~Cygni 
profiles is that they arise in the intermediate temperature region 
of the wind from an accretion disk viewed in a relatively edge-on 
manner. \cir\ then becomes an X-ray binary analog of a Broad Absorption 
Line quasar. 
At smaller radii, wind and coronal material
are likely to be heated to the Compton temperature and thus 
completely ionized. An appealing physical possibility is that the
electron scatterer discussed in \S4.1 of Brandt et~al. (1996) is 
just this material. Similar highly ionized gas has been
invoked in some models for Broad Absorption Line quasars 
(e.g., the `hitchhiking gas' of Murray et~al. 1995).  

Despite the general attractiveness of the above picture, we 
must note a potential difficulty: for a launching radius with
$v_{\rm t}\approx v_{\rm esc}$, the quantity
$N_{\rm H,launch}=\int_{r_{\rm launch}}^{\infty}n\,dr
=n_{\rm launch} r_{\rm launch}\simgt 10^{25}$~cm$^{-2}$ 
is so large that the implied column density through the wind 
is optically thick to electron scattering [here we assume 
that the wind has a radial extent $\gg r_{\rm launch}$ and
that $n=n_{\rm launch} (r_{\rm launch}/r)^2$]. Most line 
photons attempting to traverse the wind would then be Compton 
scattered out of the line. 
Shielding of the wind from the full X-ray continuum may
alleviate this problem by allowing a significant 
reduction in $n_{\rm launch}$ (see Begelman \& McKee 1983).
Alternatively, increasing $r_{\rm launch}$ helps because
$N_{\rm H,launch}\propto r_{\rm launch}^{-1}$ when we
require $\xi\simlt 1000$~erg~cm~s$^{-1}$ and thus
$n_{\rm launch}\simgt L/(r^2_{\rm launch}~1000~{\rm erg~cm~s^{-1}})$; 
in this case significant radiation pressure driving may be needed
since $v_{\rm t}>v_{\rm esc}$. $r_{\rm launch}$ cannot be 
$\gg 10^{6}$~km as the disk probably has an outer radius 
$\simlt 3\times 10^{6}$~km (see Figure~3 of Tauris et~al. 1999). 
Finally, clumping of the wind might also help because when
clumping is present $\xi=Lf/nr^2$ where $f$ is the volume 
filling factor. 
We note that Iaria et~al. (2000) have recently found evidence
for a large ionized column density ($\sim 10^{24}$~cm$^{-2}$)
along the line of sight even when \cir\ is radiating at
high luminosity; this material may be the same that makes
the P~Cygni lines. 
We are presently examining these issues in further detail. 

While we cannot rigorously rule out the possibility that the 
X-ray P~Cygni lines arise in a wind from the companion star, 
we consider this unlikely. First of all, the large wind 
velocity ($\pm 2000$~km~s$^{-1}$) implied by the profiles 
would require a high-mass secondary for the system 
(for an overview see \S2.7 of Lamers \& Cassinelli 1999),
provided the presence of the radiating compact object does not 
lead to a substantially faster wind from the companion
than would otherwise be expected. 
However, as mentioned in \S1, the bulk of the evidence suggests a 
low-mass X-ray binary nature. Furthermore, the large X-ray luminosity of
\cir\ should completely ionize an O~star wind out to 
$\simgt 5\times 10^6$~km (compare with Boroson et~al. 1999), 
while the observed P~Cygni profiles suggest that we are seeing 
the acceleration region of the outflow. 

% A future paper will present detailed spectral and 
% timing analyses of the X-ray P~Cygni profiles, 
% a complete set of the spectral features observed 
% (e.g., weaker fluorescent emission lines), and 
% fitting of the spectral continuum. 
% %
% We also intend to make additional HETGS observations to study how 
% the P~Cygni profiles change with orbital phase and luminosity;
% this should further clarify their origin.  

To our knowledge, these are the first reported X-ray P~Cygni profiles 
from an X-ray binary. Hopefully X-ray P~Cygni profiles 
will be identified and studied in other systems to provide 
geometrical and physical insight into the flows of material near
Galactic compact objects. 

% --------------------------------------------------------------------------------------

\acknowledgments
We thank all the members of the \chandra\ team for their enormous efforts, 
and we thank 
J. Chiang, 
A.C. Fabian, 
S.C. Gallagher, 
S. Kaspi,
R.A. Wade, and
an anonymous referee 
for helpful discussions. 
We gratefully acknowledge the financial support of 
CXC grant GO0-1041X (WNB, NSS), 
the Alfred P. Sloan Foundation (WNB), and
Smithsonian Astrophysical Observatory contract SV1-61010 for the CXC (NSS). 

% --------------------------------------------------------------------------------------

\newpage

%\begin{deluxetable}{lccc}
%\tablecolumns{4}
%\tabletypesize{\small}
%\tabletypesize{\scriptsize}
%\tablewidth{0pt}
%\tablecaption{P~Cygni lines from \cir
%\label{colden} }
%\tablehead{

\vbox{
\footnotesize
\begin{center}
{\sc TABLE 1\\
X-ray P~Cygni lines from \cir}
\vskip 4pt
\begin{tabular}{llccc}
\hline
\hline
{Ion} &
{Isoelectronic seq.} &
{Predicted $\lambda$} &
{Measured $\lambda$} &
{$\Delta\lambda$} \\
{Name} &
{and transition$^{\rm a}$} &
{(\AA)$^{\rm b}$} &
{(\AA)$^{\rm c}$} &
{(\AA)$^{\rm d}$} \\
\hline
\ion{Fe}{26}  & H-like (Ly$\alpha$) &  1.78 & 1.78 & 0.02 \\
\ion{Fe}{25}  & He-like             &  1.85 & 1.85 & 0.02 \\
\ion{S}{16}   & H-like (Ly$\alpha$) &  4.73 & 4.72 & 0.03 \\
\ion{S}{15}   & He-like             &  5.04 & 5.04 & 0.03 \\
\ion{Si}{14}  & H-like (Ly$\beta$)  &  5.22 & 5.21 & 0.03 \\
\ion{Si}{14}  & H-like (Ly$\alpha$) &  6.18 & 6.17 & 0.03 \\
\ion{Mg}{12}  & H-like (Ly$\gamma$) &  6.74 & 6.73 & 0.04 \\
\ion{Mg}{12}  & H-like (Ly$\alpha$) &  8.42 & 8.40 & 0.06 \\
\ion{Ne}{10}  & H-like (Ly$\delta$) &  9.48 & 9.48 & 0.09 \\
\ion{Fe}{24}  & Li-like             & 10.63 &10.63 & 0.10 \\
\ion{Ne}{10}  & H-like (Ly$\alpha$) & 12.13 &12.11 & 0.06 \\
%
% \ion{Mg}{11}  &  7.85 & 7.86 & 0.04 \\
%
\hline
\vspace*{0.02in}
\end{tabular}
\parbox{3.2in}{
% \parbox{2.5in}{
\small\baselineskip 9pt
\footnotesize
\indent
$\rm ^a${We list the transition here if it can be written compactly; for the other 
transitions see Tables~1 and 3 of Mewe et~al. (1985).} \\
$\rm ^b${From Mewe et~al. (1985).} \\
$\rm ^c${Measured intersection of the P~Cygni profile (between the emission maximum 
and absorption minimum) with a local continuum fit. The measurement accuracy 
corresponds to about 1 data bin (i.e., $\pm 0.0025$~\AA).} \\
$\rm ^d${Measured `peak-to-valley' distances for the P~Cygni profiles.}
}
\end{center}
\setcounter{table}{1}
\normalsize
\centerline{}
}

\newpage

\begin{figure}[t!]
\hbox{
\hspace*{0.6 in} \psfig{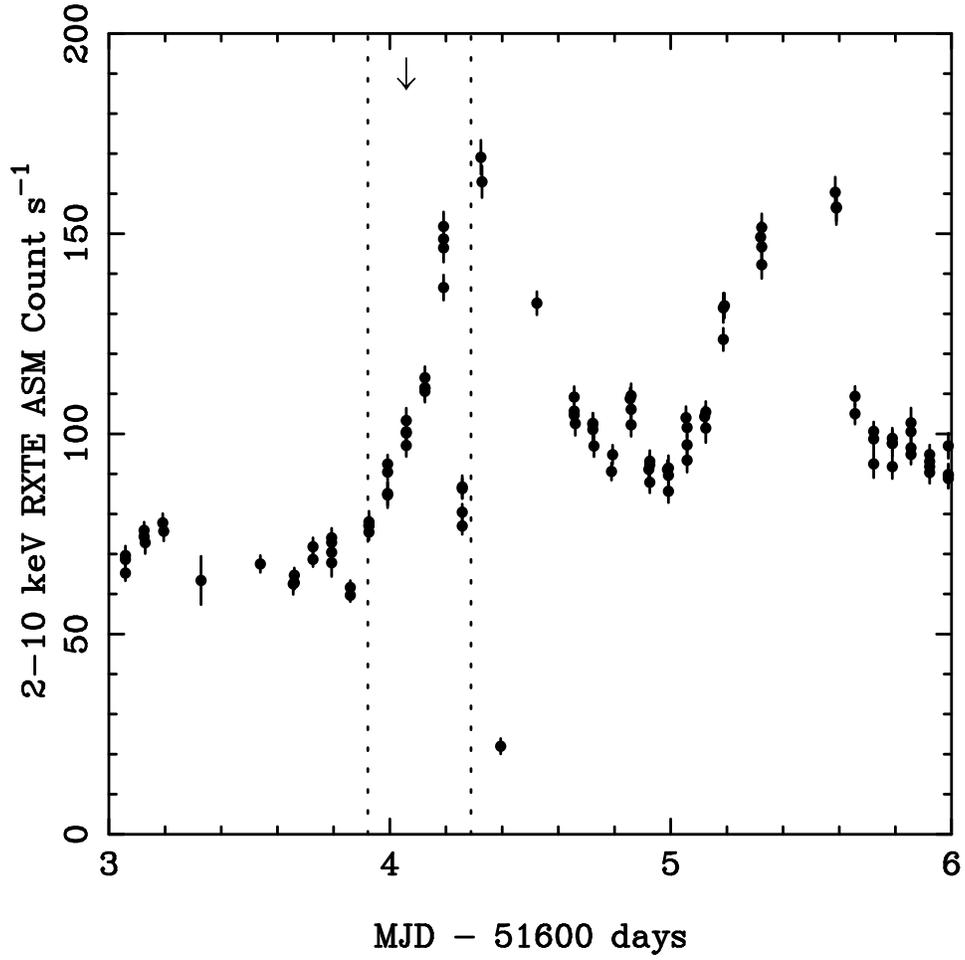}
}
\caption{
\rxte\ All-Sky Monitor (ASM) light curve of \cir\ 
around the time of our \chandra\ observation. The 
start and stop times of our \chandra\ observation are denoted 
by dotted vertical lines. The downward pointing arrow shows the 
zero-phase time of \cir. For comparison, the count rate of the 
Crab in the \rxte\ ASM is $\approx 75$~count~s$^{-1}$. 
}
\end{figure}

\begin{figure}[t!]
\hbox{
\hspace*{0.6 in} \psfig{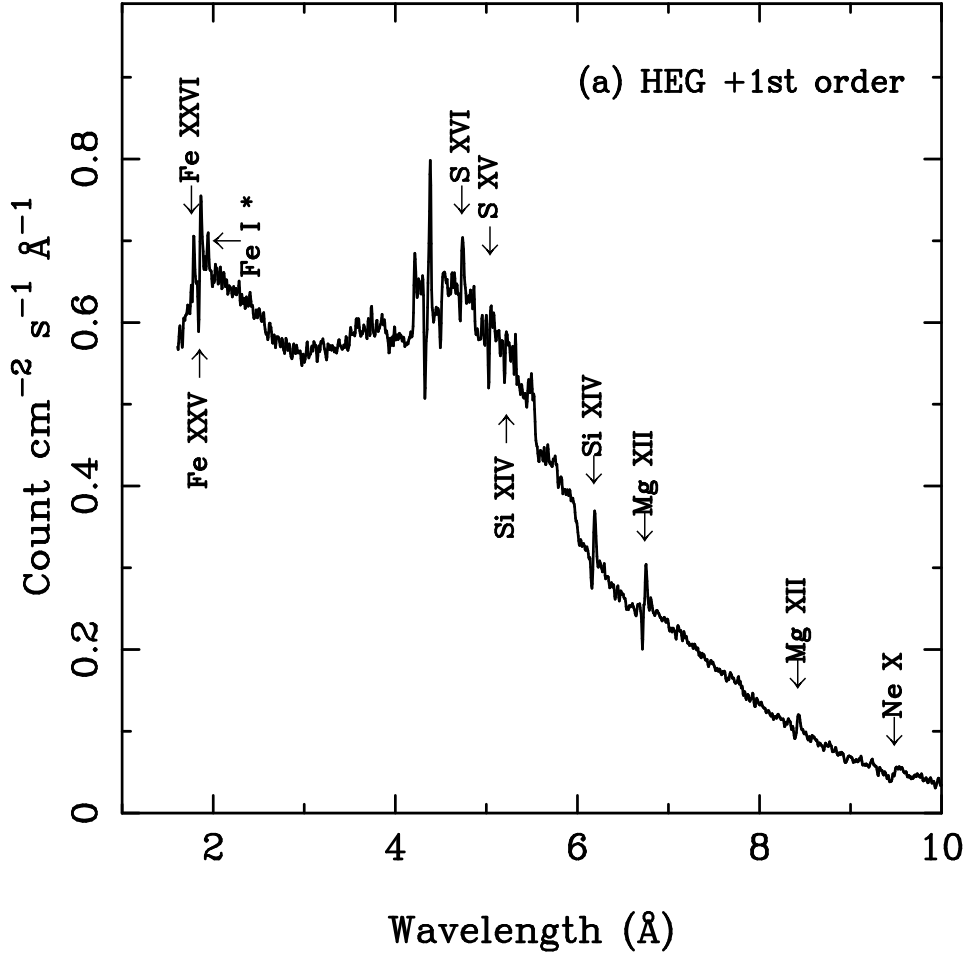}
}
\caption{
Exposure corrected (a) HEG \,$+1$st order and (b) MEG \,$-1$st order
spectra of \cir\ with a spectral binning of 0.005~\AA. These spectra have been 
smoothed with a `boxcar' filter of width 3 bins, and they have not been corrected 
for the effects of interstellar absorption. We have marked 
all the P~Cygni type spectral lines that are detected with 
$>5\sigma$ significance and that are mostly free from contamination effects
(we do not detect a P~Cygni profile from Fe~{\sc i} so this line is starred). 
The feature at 4.31~\AA\ corresponds to S~XV, but it is not marked because
its amplitude is strongly affected by uncertainties in the CCD gap correction.
}
\end{figure}

\begin{figure}[t!]
\hbox{
\hspace*{0.6 in} \psfig{figure=fig2b.ps,height=5.0truein,width=5.0truein,angle=0.}
}
% \caption{}
\end{figure}

\begin{figure}[t!]
\hbox{
\hspace*{1 in} \psfig{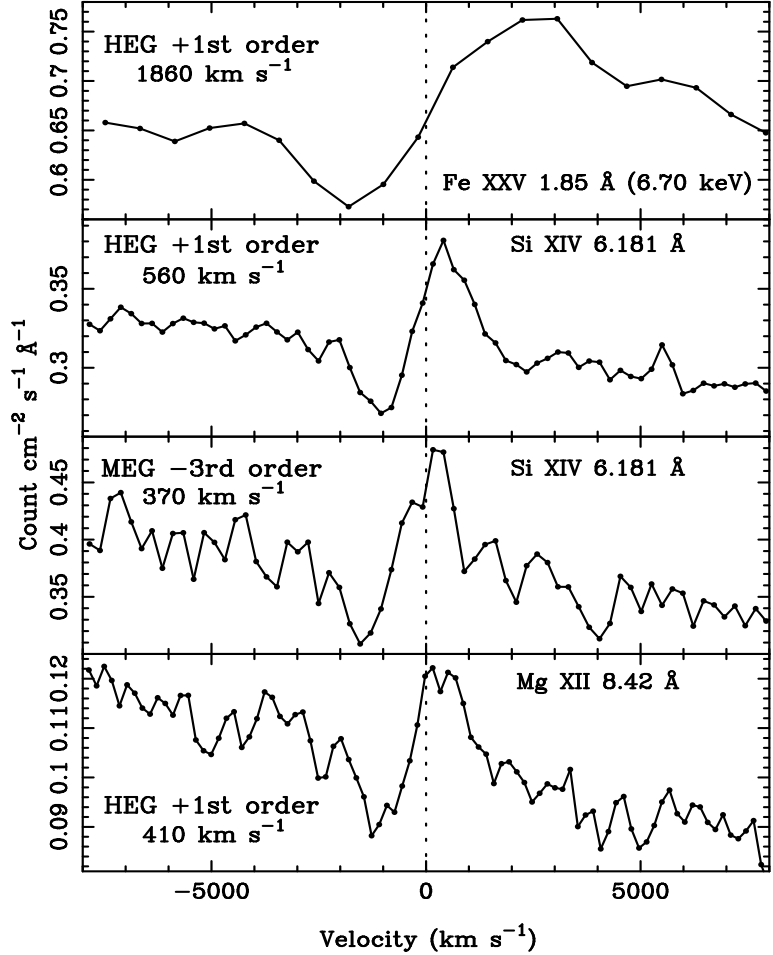}
}
\caption{
Velocity spectra showing the details of a few of the strongest 
X-ray P~Cygni profiles seen from \cir. We show the independent measurements 
of the Si~{\sc xiv} line from both the HEG \,$+1$st order and MEG \,$-3$rd order 
spectra. Typical bins in this figure have 200--1200 counts, and these spectra
have not been smoothed.  
We list the relevant velocity resolution in each panel. The lines are 
clearly broader than the instrumental resolution with velocities of 
$\pm 2000$~km~s$^{-1}$, although the Fe~{\sc xxv} line is broadened by 
the instrument. 
We have taken zero velocity to correspond to the laboratory rest wavelength,
since the radial velocity of \cir\ is not well established 
(Johnston et~al. 1999 and H.M. Johnston 2000, private communication). 
}
\end{figure}

% \vspace*{0.2in}

% --------------------------------------------------------------------------------------


\begin{thebibliography}{}

\bibitem[]{} 
Begelman, M.C., McKee, C.F. \& Shields, G.A. 1983, ApJ, 271, 70

\bibitem[]{} 
Begelman, M.C. \& McKee, C.F. 1983, ApJ, 271, 89

\bibitem[]{} 
Boroson, B., Kallman, T., McCray, R., Vrtilek, S. \& Raymond, J. 
1999, ApJ, 519, 191

\bibitem[]{} 
Brandt, W.N. \& Podsiadlowski, Ph. 1995, MNRAS, 274, 461

\bibitem[]{} 
Brandt, W.N., Fabian, A.C., Dotani, T., Nagase, F., Inoue, H., 
Kotani, T. \& Segawa, Y. 1996, MNRAS, 283, 1071

\bibitem[]{} 
Case, G.L. \& Bhattacharya, D. 1998, ApJ, 504, 761
 
\bibitem[]{} 
Fender, R., Spencer, R., Tzioumis, T., Wu, K., van der Klis, M., 
van Paradijs, J. \& Johnston, H. 1998, ApJ, 506, L121 

\bibitem[]{} 
Frank, J., King, A. \& Raine, D. 
1992, 
Accretion Power in Astrophysics
(Cambridge Univ. Press, Cambridge)

\bibitem[]{} 
Glass, I. 1994, MNRAS, 268, 742

\bibitem[]{} 
Iaria, R., Burderi, L., Di~Salvo, T., La~Barbera, A. \& Robba, N. R. 
2000, ApJ, in press (astro-ph/0009183)

\bibitem[]{} 
Inoue, H. 1989, 
in Proc. 23rd ESLAB Symp. on Two Topics in X-ray Astronomy, 
ed. N.E. White, J. Hunt \& B. Battrick
(Paris: ESA Publications), 109

\bibitem[]{} 
Johnston, H.M., Fender, R. \& Wu, K. 1999, MNRAS, 308, 415

\bibitem[]{} 
Kallman, T.R. \& McCray, R. 1982, ApJS, 50, 263

\bibitem[]{} 
Lamers, H.J.G.L.M. \& Cassinelli, J.P.
1999, 
Introduction to Stellar Winds
(Cambridge Univ. Press, Cambridge)

\bibitem[]{} 
Mewe, R., Gronenschild, E.H.B.M. \& van den Oord, G.H.J.
1985, A\&AS, 62, 197 

\bibitem[]{} 
Mignani, R., Caraveo, P.A. \& Bignami, G.F. 1997, A\&A, 323, 797

\bibitem[]{} 
Morrison, R. \& McCammon, D. 
1983, ApJ, 270, 119

\bibitem[]{} 
Murray, N., Chiang, J., Grossman, S.A. \& Voit, G.M. 
1995, ApJ, 451, 498

\bibitem[]{} 
Predehl, P. \& Schmitt, J.H.M.M. 1995, A\&A, 293, 889

\bibitem[]{} 
Raymond, J.C. 1993, ApJ, 412, 267

\bibitem[]{} 
Shirey, R.E., Bradt, H.V. \& Levine, A.M. 1999, ApJ, 517, 472

\bibitem[]{} 
Stewart, R.T., Nelson, G.J., Penninx, W., Kitamoto, S., Miyamoto, S. \& Nicholson, G.D. 
1991, MNRAS, 253, 212

\bibitem[]{} 
Stewart, R.T., Caswell, J.L., Haynes, R.F. \& Nelson, G.J., 1993, MNRAS, 261, 593

\bibitem[]{} 
Tauris, T.M., Fender, R.P., van den Heuvel, E.P.J., Johnston, H.M. \& Wu, K. 
1999, MNRAS, 310, 1165

\bibitem[]{} 
Tennant, A.F., Fabian, A.C. \& Shafer, R.A. 1986, MNRAS, 221, 27P

% \bibitem[]{} 
% Vrtilek, S.D., McClintock, J.E., Seward, F.D., Kahn, S.M. \& Wargelin, B.J. 
% 1991, ApJS, 76, 1127

\bibitem[]{} 
Weisskopf, M.C., Tananbaum, H.D., Van Speybroeck, L.P. \& O'Dell, S.L.
2000, Proc. SPIE, in press (astro-ph/0004127)

\bibitem[]{} 
Whelan, J.A.J., et~al. 1977, MNRAS, 181, 259

% \bibitem[]{} 
% Wickramasinghe, D.T., Bicknell, G.V. \& Ferrario, L. 
% 1997, 
% Accretion Phenomena and Related Outflows
% (ASP Press, San Francisco)

\end{thebibliography}
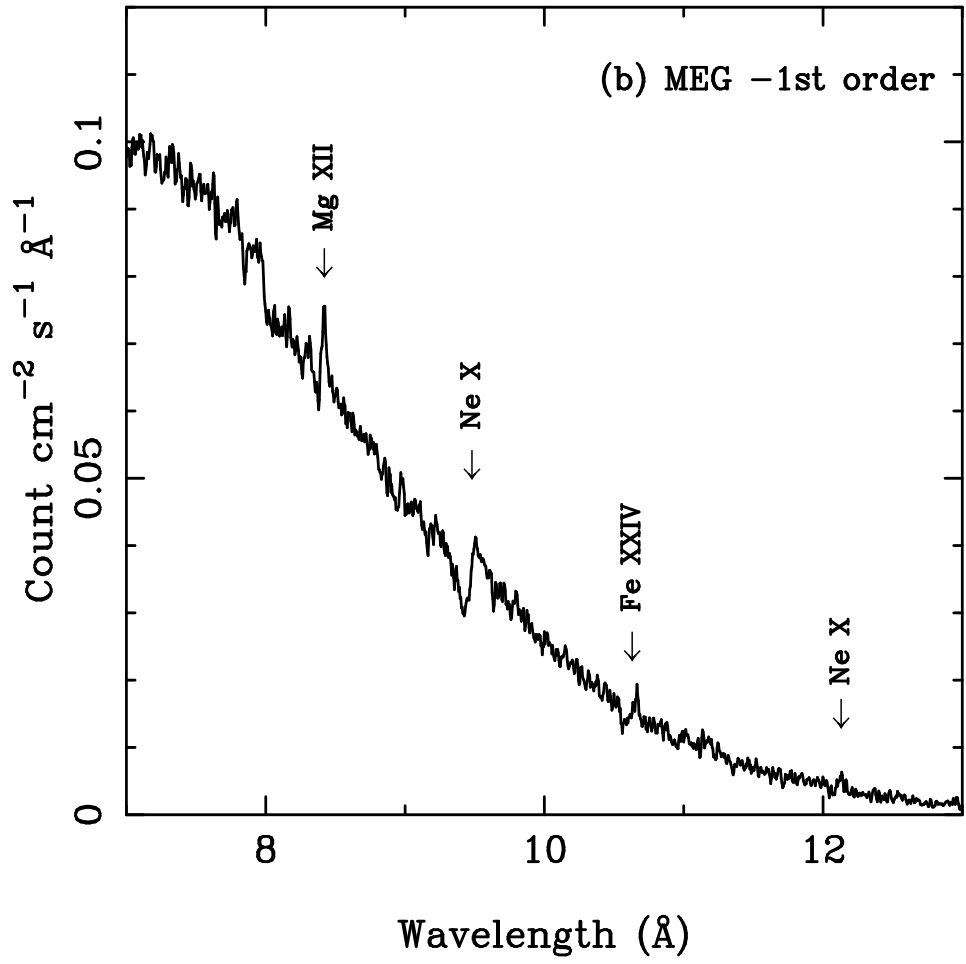
\end{document}